\documentclass[12pt,preprint]{aastex}

\newcommand\LL{\mbox{$\:\lambda\lambda $}}
\newcommand\lam{$\lambda$}
\newcommand{\kms}{km~s$^{-1}$}
\newcommand\etal{et~al.}

\newcommand\NI{N\,{\sc i}}
\newcommand\CII{C\,{\sc ii}}
\newcommand\CIII{C\,{\sc iii}}
\newcommand\OI{O\,{\sc i}}
\newcommand\OVI{O\,{\sc vi}}
\newcommand\OVIII{O\,{\sc viii}}
\newcommand\FeX{Fe\,{\sc X}}

\newcommand\FeXXIV{Fe\,{\sc xxiv}}
\newcommand\FeXVII{Fe\,{\sc xvii}}

\begin{document}

\title{FUSE Observations of Cooling Flow Gas
     in the Galaxy Clusters A1795 and A2597}

\author{W. R. Oegerle\altaffilmark{1}, 
L. Cowie\altaffilmark{2},
A. Davidsen\altaffilmark{3}
E. Hu\altaffilmark{2},
J. Hutchings\altaffilmark{4},
E. Murphy\altaffilmark{5}, \\
K. Sembach\altaffilmark{3}
B. Woodgate\altaffilmark{1}
}

\altaffiltext{1}{Laboratory for Astronomy and Solar Physics, Code 681,
Goddard Space Flight Center, Greenbelt, MD 20771}
\altaffiltext{2}{Institute for Astronomy, University of Hawaii, 
Honolulu, HI 96822}
\altaffiltext{3}{Department of Physics and Astronomy, Johns Hopkins University,
Baltimore, MD 21218}
\altaffiltext{4}{Dominion Astrophysical Observatory, HIA/NRC of Canada, 
Victoria, BC V9E 2E7 Canada}
\altaffiltext{5}{Department of Astronomy, University of Virginia, 
Charlottsville, VA 22903}

\begin{abstract}

We present far-ultraviolet spectroscopy of the cores of the massive
cooling flow clusters Abell 1795 and 2597 obtained with FUSE.  As the
intracluster gas cools through $3\times10^5$K, it should emit strongly
in the \OVI\ \LL1032,1038 resonance lines.  We report the detection of
\OVI\ \lam1032 emission  in A2597, with a line flux of $1.35 \pm 0.35
\times 10^{-15}$  erg cm$^{-2}$ s$^{-1}$, as well as detection of
emission from \CIII\ \lam977.  A marginal detection of \CIII\ \lam977
emission is also reported for A1795.  These observations provide
evidence for  a direct link between the hot ($10^7$ K) cooling flow gas
and the cool ($10^4$ K) gas in the optical emission line filaments. 
Assuming simple cooling flow models, the \OVI\ line flux in A2597
corresponds to a mass deposition rate of $\sim 40 M_\odot$ yr$^{-1}$
within the central 36 kpc.   Emission from \OVI\ \lam1032 was not
detected in A1795, with an upper limit of  $1.5 \times 10^{-15}$ erg
cm$^{-2}$ s$^{-1}$, corresponding to a limit on the mass cooling flow
rate  of $\dot{M}(28 {\rm ~kpc}) < 28 M_\odot$ yr$^{-1}$.    We have
considered several explanations for the lack of detection of \OVI\
emission in A1795 and the weaker than expected flux in A2597, including 
extinction by dust in the outer cluster, and quenching of thermal
conduction by magnetic fields.  We conclude that a turbulent mixing
model, with some dust extinction, could explain our \OVI\ results while
also accounting for the puzzling lack of emission by \FeXVII\ in cluster
cooling flows.

\end{abstract}


\section{Introduction}

X-ray observations have shown that a very hot ($\sim 10^7 - 10^8$ K)
diffuse intracluster medium pervades the cores of clusters of galaxies
\citep{fj82}. Without continued heating, this gas will cool with a 
characteristic cooling time of $t_c \sim 3T_8^{1/2}n_3^{-1}h_{50}t_H$,
where  $T_8$ is the gas temperature in units of $10^8$ K, $n_3$ is the
gas density in units of $10^{-3}$ cm$^{-3}$, $h_{50}$ is the Hubble
constant in units of 50 \kms Mpc$^{-1}$ and $t_H$ is the Hubble time
\citep{fn77}.  Consequently, if gas density  is high enough and the
temperature low enough,  the characteristic cooling time will be less
than a Hubble time \citep{cb77, fn77}. This cooling gas should ``flow''
quasi-hydrostatically onto the cD/D galaxy at the bottom of the cluster
potential well.  Typical mass flows inferred from the x-ray data are $>
100$ M$_\odot$ yr$^{-1}$ within the central $\sim 100$ kpc of the
cluster.

As the gas cools, it will emit strongly in various X-ray lines of highly
ionized Fe and O for $T \la 10^7$ K \citep{canizares88}.  Emission lines
of \OVIII\ Ly$\alpha$ and \FeXXIV\/ arising in this hot gas have been
detected in the cores of clusters \citep{peterson01, tamura01}.  Spatially
resolved x-ray spectroscopy has confirmed the presence of distributed
cooling gas in clusters \citep{af97}.
As the gas cools rapidly through the $T \sim 3 \times 10^5$ K regime, it
should emit strongly in certain UV resonance lines -- particularly
the \OVI\ \LL1032,1038 doublet \citep{ec86, voit94}. Finally, as the
gas cools to $T \sim 10^4$ K, it should emit strongly in the hydrogen
lines (H$\alpha$, Ly$\alpha$).  Optical H$\alpha$ emission from cooling
flows has been detected toward several clusters \citep{heckman81,
cowie83, hu85},   \citet{hu92} has also reported extended Ly$\alpha$
emission from a number of clusters having H$\alpha$ emission.

Emission lines from cooling flow gas at temperatures $\sim 1-3 \times
10^5$ K have yet to be convincingly detected. \citet{dixon96} observed
several clusters, including A1795, with the FUV spectrograph on the {\em
Hopkins Ultraviolet Telescope} (HUT) in an attempt to detect emission
from \OVI\/.  No detections were made, although in the case of the
strongest X-ray cooling flow cluster observed, A1795, the HUT exposure
time was only 2000 sec.  In efforts to detect gas at temperatures of
$10^{5.5} - 10^{6.5}$K, a number of investigators have attempted to
detect  the \FeX\ \lam6374 coronal line in cooling flow clusters
\citep{anton91, megan94, yan95}.  \citet{yan95} provided the the most
sensitive search reported to date, and found no evidence for the line
and set an upper limit for the surface brightness of $10^{-18}$ erg
cm$^{-2}$ s$^{-1}$ arcsec$^{-2}$.   Emission lines from \FeXVII\ arising
in gas  at $T \la 5 \times 10^6$K have also not been detected in a
recent {\em XMM} observation of A1795 \citep{tamura01}.

Detection of emission from gas at $\sim 3 \times 10^5$ K would 
provide the lynchpin to
physically connect the previously detected hot and cool gas components
in the cores of clusters.  We have therefore begun a program with 
the {\em Far Ultraviolet Spectroscopic Explorer} (FUSE)
to search for emission from \OVI\ \LL1032,1038 in the cores of the
x-ray bright, cooling flow clusters A1795, A2199 and A2597.  Results for
A1795 and A2597 are presented in this paper. FUSE has twice the effective
area of HUT at \OVI\ \lam1032, which, when coupled with longer
integration times and higher spectral resolution can provide a 
considerable increase in sensitivity over the HUT.  For example, the
FUSE observations reported here for A1795 are a factor of 15 more sensitive
than the HUT observations.

Throughout this paper, we have used a value of $H_\circ =  50$ \kms\
Mpc$^{-1}$.  All distances scale as the inverse of
$H_\circ$, while observed line luminosities and derived cooling flow
mass deposition rates scale as $H_\circ^{-2}$.

\section{FUSE Observations}

Far-ultraviolet spectra of the cores of A1795 and A2597 have been obtained
with the FUSE observatory.  A description of the FUSE
instrument and its in-orbit performance is described in papers by
\citet{moos00} and \citet{sahnow00}.  In brief, FUSE spectra cover the
wavelength range $905-1185$\AA\ with a spectral resolution of $\sim
15-20$ \kms\ (depending on wavelength) for point sources. 
FUSE spectra are recorded in four separate optical channels, two of
which employ LiF optics covering the wavelength range $\sim 1000 - 1185$\AA\
and two which use SiC optics covering the $905-1090$\AA\ range.  Dispersed
light is focused onto two photon-counting microchannel plate detectors.

All observations were carried out with the LWRS spectrograph entrance
aperture ($30'' \times 30''$). For extended emission which fills the
LWRS aperture, the instrumental line width is $\sim 100$ \kms\/, which
is narrower than the expected line width of cooling flow gas in the core
of a cluster of galaxies.

The data were processed with the FUSE calibration software ({\it
calfuse}, version 1.8.7).  Since the signal from the source is expected
to be weak, extra care was taken in subtraction of the detector
background.  Photons events with pulse heights less than 4 or greater
than 15 were screened out of the data, since these events are
predominately background in nature. Furthermore, the standard background
subtraction employed by {\it calfuse}, which is adequate for bright
objects, was not used.  Instead, the background level on each detector 
was determined from a co-addition of all exposures for each cluster. 
The local background in the vicinity of the expected line positions was
measured on the 2D spectral image at locations just above and below the
spectra, and were then subtracted from the extracted spectra.  

We have searched the extracted spectra for emission from \OVI\ \lam1032
and \CIII\ \lam977.  \OVI\ \lam1032 is redshifted to 1097\AA\ and
1117\AA\ in A1795 and A2597 respectively.  These wavelengths fall  in
the LiF channels for both detectors 1 and 2, and for the SiC channel in
detector 2 (for A1795 only),  and is outside the wavelength range of the
SiC channel in detector 1. The combined effective area of the 2 LiF
channels at $\sim 1100$\AA\/,  based on in-flight calibrations, is about
52 cm$^{-2}$.   The SiC2 channel at this wavelength has an effective
area of only $\sim 3$  cm$^{-2}$, and hence we have not used the SiC2
spectrum in our analysis.  The \CIII\ \lam977 line is also redshifted to
a wavelength covered by both LiF and SiC channels. However, the combined
effective areas of the LiF channels is much greater than the SiC
channels, and given the fact that the signal is close to the detector
background, we have used only the LiF spectra from channels 1 and 2.

Terrestrial airglow lines in the far-UV \citep{feldman00} have also
posed a special problem in our search for emission lines from cooling
flows.    For example, in A1795 the \CIII\ \lam977 line is redshifted to
1038.7\AA\/,  which is adjacent to an \OI\ airglow line. In A2597,
the \CIII\ line is redshifted to 1057.4\AA\/, which is potentially
contaminated by a terrestrial N$_2$ emission band.  Finally, in A1795,
the \OVI\ \lam1032 line is redshifted to coincide with a  band of \NI\
airglow lines. We have thoroughly investigated all these possible
contaminations and discuss the results below.

\subsection{A1795}

A1795 was observed on February 23, 2000 with an integration time of
37,189 seconds.  The cluster has a redshift of 0.0631 \citep{oh94},
hence the FUSE aperture subtends $50 \times 50$ kpc at the cluster. The
spectrograph aperture was centered on the cD galaxy at coordinates: 
$\alpha(2000) = 13^h 48^m 52.5^s, \delta(2000)= +26^\circ 35' 35''$. 

Any \OVI\ \lam1032 emission from A1795 would be  redshifted to an
observed wavelength of 1097\AA\/, which is unfortunately at nearly the
same wavelength as a complex of \NI\ terrestrial airglow lines
\citep{feldman00}.  These lines are strong only when looking at the
illuminated earth.  However, they are also present in spectra taken near
the earth limb, and are faintly visible throughout the daylit portion of
the  orbit.  To avoid the effects of these airglow lines as much as
possible, we have restricted our analysis of the \OVI\ line  to data
obtained during spacecraft orbital night, limiting the integration time
to 17 ksec. 

The LiF1 and LiF2 spectra of A1795 were coadded
and the resulting spectrum at the expected
wavelengths of the redshifted \OVI\ lines is shown in Figure 1, where
the data has been binned by 0.2\AA\ ($\sim 60$ \kms).  A faint continuum
from the cD elliptical galaxy is detected at a level of 
$\sim 1.5 \times 10^{-15}$ erg cm$^{-2}$ s$^{-1}$ \AA$^{-1}$, in good 
agreement with the UV continuum flux reported by \citep{hu92} at wavelengths 
$\lambda > 1215$\AA\ from an IUE large aperture observation.
This relatively large far-UV flux from an old 
elliptical galaxy at a redshift of 0.0631
indicates the presence of substantial star formation activity. 
\citet{hu88} has estimated an ongoing star formation rate of $\sim 1$
M$_\odot$ yr$^{-1}$ from the IUE data.  \citet{smith97} and \cite{mittaz01}
have reported broad-band UV emission from the center of A1795, and 
concluded that a population of early type stars is present, which
presumably formed from the cooling flow.

In Figure 1, we note possible Galactic absorption lines of H$_2$ at
1094 and 1100\AA\/, which makes the spectrum appear somewhat
noisier than it truly is.  These absorption features do not correspond
to any metal absorption lines at the redshift of A1795.  
In any event, there is no obvious \OVI\
emission. After subtracting a continuum fit to the galaxy spectrum, we
compute a flux in the \lam1032 line of $6 \pm 6 \times 10^{-16}$ erg
cm$^{-2}$ s$^{-1}$,  where we have integrated from $-300$ to $+300$
\kms\/ about the expected line center. The uncertainty quoted here is
purely a random error.  There is also a systematic uncertainty in the
continuum placement, so we  estimate a 2$\sigma$ upper limit to the flux
in \lam1032 line of $1.5 \times 10^{-15}$ erg cm$^{-2}$ s$^{-1}$, or in
terms of the average surface  brightness within the FUSE LWRS aperture,
$\mu$(\OVI\/)$ < 1.7 \times 10^{-18}$ erg cm$^{-2}$ s$^{-1}$
arcsec$^{-2}$ or equivalently, $7 \times 10^{-8}$ erg cm$^{-2}$ s$^{-1}$
sr$^{-1}$.   This then yields an upper limit to the luminosity in the
\OVI\ \lam1032 line of   $L_{obs}($\OVI\/$) = 2.5 \times 10^{40}$
erg~s$^{-1}$. 

We have have also searched the A1795 spectrum for \CIII\ \lam977
emission using the full 37 ksec integration taken during both  orbital
night and day. In Figure 2, we display the $1020-1060$\AA\ spectral region
from that dataset. The background subtraction appears to be
quite good, since the \CII\ \lam1036 absorption line from the Galactic
interstellar medium is black at line center, as expected.     A faint
emission line appears to be present adjacent to the \OI\ \lam1039.2
terrestrial airglow line, and we suggest that this is emission from
\CIII\ \lam977 in the cooling flow of A1795, redshifted to 1038.7\AA\/. 
The line flux is uncertain due to the airglow contamination, but we
estimate a flux of $1.5\pm0.7 \times 10^{-15}$ erg cm$^{-2}$ s$^{-1}$;
ie. only a $\sim 2\sigma$ detection. We also note broad absorption extending
from $\sim 1032$\AA\ to 1037\AA\/ in the observed frame.   Absorption at
the  two ends of this range are  due to \OVI\ \lam1032 and \CII\
\lam1036 from the Galactic ISM.  The absorption at 1034\AA\ corresponds
in wavelength to redshifted Ly$\gamma$ in A1795.  Absorption by neutral
hydrogen in A1795 could, in principle, be confirmed by detecting  
Ly$\beta$ absorption.  However, Ly$\beta$ is redshifted to 1090\AA\/,
which is very near the edge of segment B of detector 1, where 
calibration is problematical.  This wavelength is not recorded on
detector 2, since it lies in the gap between segments A and B.

\subsection{A2597}

A2597 was observed on June 26, 2000 for a total integration time of
43,032 seconds.  The spectrograph aperture was centered on the cD
galaxy at coordinates: $\alpha(2000) = 23^h 25^m 19.7^s, \delta(2000)=
-12^\circ 07' 27''$. The redshift of the A2597 cD galaxy is 0.0823 
\citep{heckman89,taylor99}, thereby shifting the \OVI\ \lam1032 line to
to 1116.86\AA\ in the observed frame.   This is in a region of the
spectrum where there is no observed or expected  airglow emission
\citep{feldman00}.  We have therefore used spectra obtained during both
orbital day and night in our analysis of A2597.  The co-added spectrum
from the LiF1 and LiF2 channels is shown in Figure 3, binned to
0.2\AA\/.  A weak, but clearly detected emission line is evident at the
expected position of redshifted \OVI\ \lam1032.  The emission line is
clearly visible in both the individual LiF1 and LiF2 spectra, so it is
very unlikely to be due to a detector artifact.  Indeed,  no such
emission feature is observed at 1117\AA\ in the combined day and night
(34 ksec) observation of A1795.  We therefore conclude that this weak
emission line is redshifted \OVI\ \lam1032 from the cooling flow nebula
of A2597.

We measure a flux in the \lam1032 line of $1.3 \pm 0.35 \times 10^{-15}$
erg cm$^{-2}$ s$^{-1}$,  where we have integrated from $-300$ to $+300$
\kms\/ about the expected line center.  The FUSE aperture subtends $65
\times 65$ kpc at the cluster, resulting in an average
surface brightness within the FUSE LWRS aperture of $\mu$(\OVI\/)$ = 1.4
\times 10^{-18}$ erg cm$^{-2}$ s$^{-1}$ arcsec$^{-2}$ or equivalently,
$6 \times 10^{-8}$ erg cm$^{-2}$ s$^{-1}$ sr$^{-1}$. The luminosity in
the \OVI\ \lam1032 line is then $L_{obs}($\OVI\/$) = 3.6 \times 10^{40}$
erg~s$^{-1}$.

An examination of the spectrum has also revealed an emission line at 
1057.4\AA\/, the
expected position of redshifted \CIII\ \lam977 (see Figure 4).  
\citet{feldman00} have observed emission from terrestrial N$_2$ at this
wavelength when pointed at the bright earth.  In order to determine if
the observed line in our data is terrestrial in origin, we have also
extracted spectra of A2597 obtained during orbital night.  Although
the night-only data was restricted to 14 ksec, the emission line at
1057\AA\ is detected quite easily.  Furthermore,  the flux in the
line is the same during the night and daylit portions of the
orbit.  Finally, we note that there is no emission line at this wavelength
present in the A1795 spectra.  We conclude that the evidence is strong 
that the emission line at 1057\AA\ is not of terrestrial
origin, and is in fact redshifted \CIII\ \lam977 from A2597.
The measured line flux is $4.9 \pm 0.7 {\rm (random)} 
\pm 1.0 {\rm (systematic)} \times 10^{-15}$
erg cm$^{-2}$ s$^{-1}$ measured within $\pm 300$ \kms of the expected
line center.

\section{Discussion}

The flux in the \OVI\ line is a function of the temperature of the gas,
its distribution in the cluster, and the mass flow rate in the cooling
flow. \citet{ec86} and \citet{voit94} have computed the expected
luminosities of ultraviolet emission lines from a gas cooling from a
temperature $>10^6$K to $10^4$K.  The luminosity in the \OVI\ line is
given by

\[
L({\rm OVI}) = \frac{k\dot{M}}{\mu m_H} A_O \int_{T_{min}}^{T_{max}}
  \left( \frac{3}{2} + s \right)  \chi \frac{f_{i}j({\rm OVI})}{\Lambda}dT
\]
where $\dot{M}$ is the mass flow rate, $\mu m_H$ is the mean mass per atom,
$A_O$ is the abundance of oxygen, $f_{i}$ is the fraction of oxygen in the
O$^{+5}$ ionization stage, $\chi$ is the number of particles per H atom,
$j($\OVI\/) is the volume emissivity in the \OVI\ line, and 
$\Lambda$ is the radiative
cooling function \citep{cowie81,ec86}.  The gas cools from T$_{max}$ 
down to T$_{min}$ with an evolution 
that could be isobaric, isochoric or some combination of
the two; $s$ is 1 for isobaric cooling and 0 for isochoric.  
The intracluster medium (ICM) has abundances that are typically
$0.1-1$ times solar \citep{mushot92, af98} which can be understood
in terms of the ICM being composed of 
galactic ejecta mixed with primordial gas.  Note in the
above equation that lower metallicity (of oxygen in the present case)
is offset by a similar decrease in the cooling function,
assuming that the abundance of oxygen is not dissimilar to other metals
providing the bulk of the cooling. Hence, the line luminosity is rather
insensitive to metallicity, as first pointed out by
citet{cowie81}.  We also note that the assumption made here is that
this is the luminosity of an optically thin, extended region, with no
self-absorption.  \citet{ec86} have calculated UV line luminosities for solar
abundances, with T$_{min} = 10^4$~K and T$_{max} = 10^6$~K.  \citet{voit94}
arrive at very similar results.
Adopting an intermediate case between isobaric and isochoric cooling,
the predicted luminosity in the \OVI\ \lam1032 line is $\sim 9 \times
10^{38}\dot{M}_\odot$ erg~s$^{-1}$, where $\dot{M}_\odot$  is the mass
cooling flow rate in solar masses per year.   For an optically thin
plasma, the luminosity of the 1038\AA\ line is half of that of the
1032\AA\ line.

\subsection{A1795}

Based on model prediction above, our upper limit to the observed
luminosity of A1795 in the \OVI\ \lam1032 line  implies an upper limit
to the mass cooling flow rate of only $\sim 28$ M$_\odot$ yr$^{-1}$.  

\citet{edge92} determined the mass flow rates in a large sample of
clusters observed with the {\em Einstein Observatory} and EXOSAT, using
a  deprojection analysis of the x-ray images.  For A1795, they found a
mass deposition rate of $\dot{M} = 478M_\odot$ yr$^{-1}$ within a radius
of $230$ kpc.  \citet{allen00} computed mass flow rates for a  sample of
30 of the most luminous X-ray clusters, by applying a  similar
deprojection technique to ROSAT HRI images,  and analyzing spatially
resolved spectra of the clusters obtained with ASCA.  Within a cooling
radius of $r_{cool} = 188$ kpc,  \citep{allen00} finds a spectroscopic
mass flow rate of $\dot{M}_S =  300M_\odot$ yr$^{-1}$. However,
deprojection of the ROSAT images yielded $\dot{M}_I =  449M_\odot$
yr$^{-1}$. A deprojection analysis including intrinsic absorption using
the same column densities determined from the spectral analysis of the
ASCA data yielded a corrected value of $\dot{M}_C = 1038M_\odot$ yr$^{-1}$.

In the generally adopted cooling flow picture, gas begins cooling from a
temperature of $\sim 5 \times 10^7$K at several hundred kpc from the
cluster center, and cools as it flows inwards with $\dot{M}(r) \propto r$, 
where $\dot{M}(r)$ is the integrated mass deposition rate interior
to $r$. For A1795,  the FUSE aperture has an effective   radius of $28$
kpc, or $0.15r_{cool}$ as determined by \citet{allen00}.  Assuming a
mass deposition rate proportional to radius, then the x-ray mass cooling
rate within the FUSE aperture is  $\dot{M}_X = 45-155M_\odot$ yr$^{-1}$
depending on whether one adopts $\dot{M}_S$, $\dot{M}_I$ or  $\dot{M}_C$
from \citep{allen00}.  Our upper limit to the mass cooling flow rate of
$28M_\odot$ yr$^{-1}$ from the  non-detection of \OVI\ \lam1032 emission
is about half of that derived from the ASCA x-ray spectra, assuming no
absorption of the \OVI\ line.

\subsection{A2597}

The observed \OVI\ \lam1032 luminosity in A2597 implies a  cooling flow
rate of $\sim 40$ M$_\odot$ yr$^{-1}$, based on the model described above.

\citet{sarazin95} has derived a cooling rate for the ambient hot gas of
$\dot{M} =  327M_\odot$ yr$^{-1}$ within a radius of $r_{cool} =  130$
kpc from ROSAT HRI x-ray images.  The effective radius of the FUSE
aperture at the redshift of A2597 corresponds to  $46$ kpc, or
$0.35r_{cool}$.  Scaling the mass deposition rate with radius, we then
find an x-ray mass cooling rate within the FUSE aperture of $\dot{M}_X =
115M_\odot$ yr$^{-1}$, which is almost 3 times the cooling rate of
$40M_\odot$ yr$^{-1}$ inferred from the the luminosity of the  \OVI\
line.

\subsection{Dust Scattering}

One possible way to diminish the observed \OVI\ emission is through
scattering by dust.  \citet{edge99} have reported the existence of
significant amounts of dust in the cores of some cooling flow clusters,
based on submillimeter observations.  \citet{hu92} has investigated the
Ly$\alpha$/H$\alpha$ emission line ratios in cooling flow clusters, and
found an average excess reddening, above the foreground extinction, of
$E(B-V)$ in the range $0.09-0.25$. \citet{hu92} discusses the location
of the dust in the cluster, and favors the idea that the bulk of the
dust is in the extended cluster component, instead of embedded in the
localized cooling flow region.  An extended dust component was found to
be in rough agreement with estimates of cluster extinction based on
background galaxy counts \citep{kara69}, and is supported by the lack of
a correlation between $E(B-V)$ with $\dot{M}$ in cooling flow systems. 
If this is the case, then regardless of the geometry of the cooling flow
region, \OVI\ emission from the core of the cluster will have to pass
through the dust in the outer cluster on its way to the observer. 

\citet{hu92} derived a value of $E(B-V) = 0.14_{-0.10}^{+0.08}$ for
A1795 (although take note of the large uncertainties, which are
primarily  systematic).  Assuming an upper limit to $E(B-V)$ of 0.22,
this translates into $<2.9$  magnitudes of extinction in the
far-ultraviolet, computed assuming a standard Galactic dust extinction curve
with $R_V = 3.1$.  If we assume that all of this extinction arises in
the outer cluster and scatters the \OVI\ emission,  then our observed,
extinction-corrected  line luminosity is $L_{corr}($\OVI\/$) < 3.6
\times 10^{41}$ erg~s$^{-1}$. This then raises the limit to the mass
cooling rate to  $\dot{M}_{OVI} < 400 M_\odot$ yr$^{-1}$, which does not
constrain the x-ray models.  For A2597, \citet{hu92} quotes $E(B-V) <
0.11^{+0.08}_{-0.09}$ which would result in an extinction corrected mass
flow rate derived from the \OVI\ observation of $\dot{M}_{OVI} \sim
150^{+140}_{-100} M_\odot$ yr$^{-1}$,  which is in reasonable agreement
with the x-ray derived mass flow rate within the FUSE aperture, given
the uncertainties in the actual extinction value.  

\citet{vd95} have pointed out that dust in the ICM may be unlike
Galactic dust, and may have different extinction properties.  Indeed, it
is clear that extinction properties are dependent upon the  past
processing history and environment in which the dust grains reside
\citep{mathis92,gordon98}.  However, extinction in the far-UV is
dominated by small grains, and these would be subject to destruction in
the hot ICM by sputtering.  In hot gas with density $n_H$ cm$^{-3}$, a
grain of size $a_\mu$ microns will be destroyed in $\sim 2
\times 10^6 a_\mu n_H^{-1}$ yr \citep{vd95}.  So, grains smaller than
$1\mu$m in size would be destroyed in less than a Hubble time unless the
grains were near the outskirts of the cluster where  $n_H < 10^{-4}$
cm$^{-3}$.   If intracluster dust at large cluster radii existed and had an
extinction law like that of the Small Magellanic Cloud (ie. containing
small grains),  then the ICM extinction could be $\sim 0.5-1$ mag
greater at \OVI\ than it would if an average Galactic law were assumed.

Hot gas near the cluster center should be free of dust grains with sizes
$<1\mu$m due to sputtering.  So, any clouds that condense out of the
cooling flow should also be dust-free.  However, the gas surrounding the
central galaxies in many cooling flow clusters appears to be dusty
\citep{sparks89,mcnamara92}. \citet{dv93} have reported significant
depletion of calcium in cooling flow emission line regions, and this is
generally attributed to dust depletion. In the core of A1795, 
\citet{pinkney96} have resolved a 7 kpc long dust lane,  whose
morphology matches that of the H$\alpha$ emission.  \citet{koekemoer99}
have also resolved an elongated dust lane southeast of the nucleus of
A2597, together with patches of dust near the cluster core.  This dusty
gas may have been stripped from a gas-rich galaxy passing through the
cluster core, or it might be the remains of a merger of such a galaxy
with the central cD galaxy.  Consequently, these dusty filaments may be
very dynamic, and may form and be destroyed continuously throughout the
life of the cluster.  The effect of dust in the core of the cluster on
\OVI\ emission would likely depend on its geometry and/or filling
factor.  If the dust were confined to optical emission line filaments
having a small filling factor, then the the observed
Ly$\alpha$/H$\alpha$ emission line ratios could be altered by this dust,
and yet the \OVI\ emission, which arises in the intercloud regions (or
at cloud interfaces), could suffer much less extinction than the optical
emission lines.

\subsection{Conduction and Mixing}

The existence of magnetic fields in intracluster gas with strengths of
order 1 $\mu$G have been inferred from observations of Faraday rotation
of radio sources seen through the ICM \citep{kim91}.  Tangled magnetic
fields, or field lines that run perpendicular to the temperature
gradient, can prevent thermal conduction in an ionized gas.  Recent
x-ray observations of two clusters, A2142 \citep{markevitch00} and A3667
\citep{vikhlinin01}, by {\em Chandra} have revealed what appears to be
central, cooler intracluster gas separated from the outer, hot gas by a
very sharp density discontinuity.  \citet{ettori00} have pointed out
that this sharp temperature discontinuity could not be maintained
without a significant suppression of heat conduction.

Recent {\em Chandra} x-ray images of the core of A1795 reveal an 80 kpc
long filament of x-ray emission extending to the SSE of the cD galaxy
\citep{fabian01a}. There is a remarkable morphological  similarity
between this x-ray emission and the H$\alpha$ emission. Although the
nature of the filament is unclear (ie. a cooling wake, or ram pressure
stripped gas from a moving galaxy, etc; see \citet{fabian01a}) there
appears to be a direct physical relationship between the x-ray cooling
gas and the optical emission.  Although not all of this filamentary
structure is within the FUSE aperture, the brightest parts of the
filament near the cD galaxy are within the aperture (see Figure 5).  If
the cool clouds in the filament are small enough,  they will be
destroyed by thermal evaporation in the hot, ambient gas, unless
conduction is quenched by tangled magnetic fields.  The existence of the
filament may then indicate that conduction is ineffective in cooling the
hot x-ray emitting gas.

The absence of \OVI\ emission in A1795, the rather weak \OVI\ emission
in A2597, and the absence of observed \FeXVII\ lines in cooling flows
\citep{fabian01b} could then be explained by the suppression of
conduction.   Nevertheless, we do observe \OVI\ emission in A2597 at a
level consistent with expectations from x-ray observations if we correct
the \OVI\ flux for a reasonable amount of scattering by dust.  We are
then left to ponder why no x-ray emission has been detected by {\em XMM} from
gas with temperatures less than $\la 4\times10^6$K.  A possible
explanation is provided by turbulent mixing of hot and cool gas
\citep{begelman90}. \citet{loewenstein90} have pointed out that the 
inner regions of cooling flows are likely to be highly turbulent, with
turbulent velocities of order $\sim 100$ \kms\/.  Assuming the presence
of entrained cool clouds, any relative motion between the hot gas and
the cool clouds will excite Kelvin-Helmholtz instabilities that will
grow and create a mixing layer at the interface.  The surface area of
the cool material is increased dramatically by this turbulence, and
allows efficient diffusion of energy between the hot and cool phases
\citep{begelman90,slavin93}, resulting in a very short mixing time. The
steady-state temperature of the mixed gas is intermediate between the
hot and cool gas with a value of $\xi\sqrt{T_cT_h}$ \citep{begelman90},
where $T_c$ and $T_h$ are, respectively, the cool and hot gas temperatures
and $\xi$ is a factor of order unity, and depends on the  ratio of the
efficiencies with which hot and cool gas deposit mass and  energy into
the mixing layer. For the case we envision in cluster cooling flows, if
cool gas at temperature $\sim 10^4$K and hot gas at temperature 
$\sim 10^7$K are mixed, then the intermediate gas temperature in the
mixing layer will be $\sim  3\times 10^5$K.  This gas is at the peak of
the cooling curve, and would rapidly cool and emit \OVI\ \lam1032, while
being too cool to emit the \FeXVII\  x-ray lines.
  
\section{Summary}

We have obtained far ultraviolet spectroscopy of the cores of the galaxy
clusters A1795 and A2597, in order to detect emission lines arising in
cooling flow gas with temperatures of $1-3\times 10^5$ K. We do not
detect \OVI\ \lam1032 in A1795, with a $2\sigma$ upper limit of  $\sim
1.5\times 10^{-15}$ erg cm$^{-2}$ s$^{-1}$.  This corresponds to an
upper limit to the mass deposition rate of $28 M_\odot$ yr$^{-1}$ within
the central 28 kpc, assuming cooling models of \citet{ec86} and
\citet{voit94}. We report a marginal ($\sim 2\sigma$) detection of
emission from \CIII\ \lam977 in A1795.  We have detected \OVI\ \lam1032
and  \CIII\ \lam977 emission in A2597 providing direct evidence linking
the cool gas at $10^4$K with the ambient hot, $10^7$K gas.  In
collisional ionization equilibrium, O$^{+5}$ and C$^{+2}$ are formed at
temperatures of $3 \times 10^5$ K and $8 \times 10^4$ K, respectively.
The observed \OVI\ \lam1032 line flux in A2597 is $\sim 1.3 \pm 0.35
\times 10^{-15}$ erg cm$^{-2}$ s$^{-1}$ and corresponds to mass
deposition rate of $\dot{M}(<36 {\rm ~kpc}) \sim  40 M_\odot$ yr$^{-1}$.
 The non-detection of \OVI\ \lam1032 in A1795 could be due to scattering
by dust, or to quenching of thermal conduction by intracluster magnetic
fields, which prevents the gas from cooling below $\sim 5\times10^6$K. A
turbulent mixing model, together with some dust extinction, appears to
adequately describe our detection of \OVI\ \lam1032 emission in A2597,
while also accounting for recent {\em XMM} spectrosocopic results, which
indicate a  lack of \FeXVII\ emission, and hence a lack of cooling flow
gas at temperatures of $\sim 4 \times 10^6$K.

\acknowledgments

This work is based on data obtained for the Guaranteed Time
Team by the NASA-CNES-CSA FUSE mission operated by the
Johns Hopkins University. Financial support to U. S.
participants has been provided by NASA contract NAS5-32985.
We thank Joel Bregman and Rich Mushotzky for informative discussions
on cooling flows and the recent {\em XMM} results.

\newpage

\begin{figure*}
\plotone{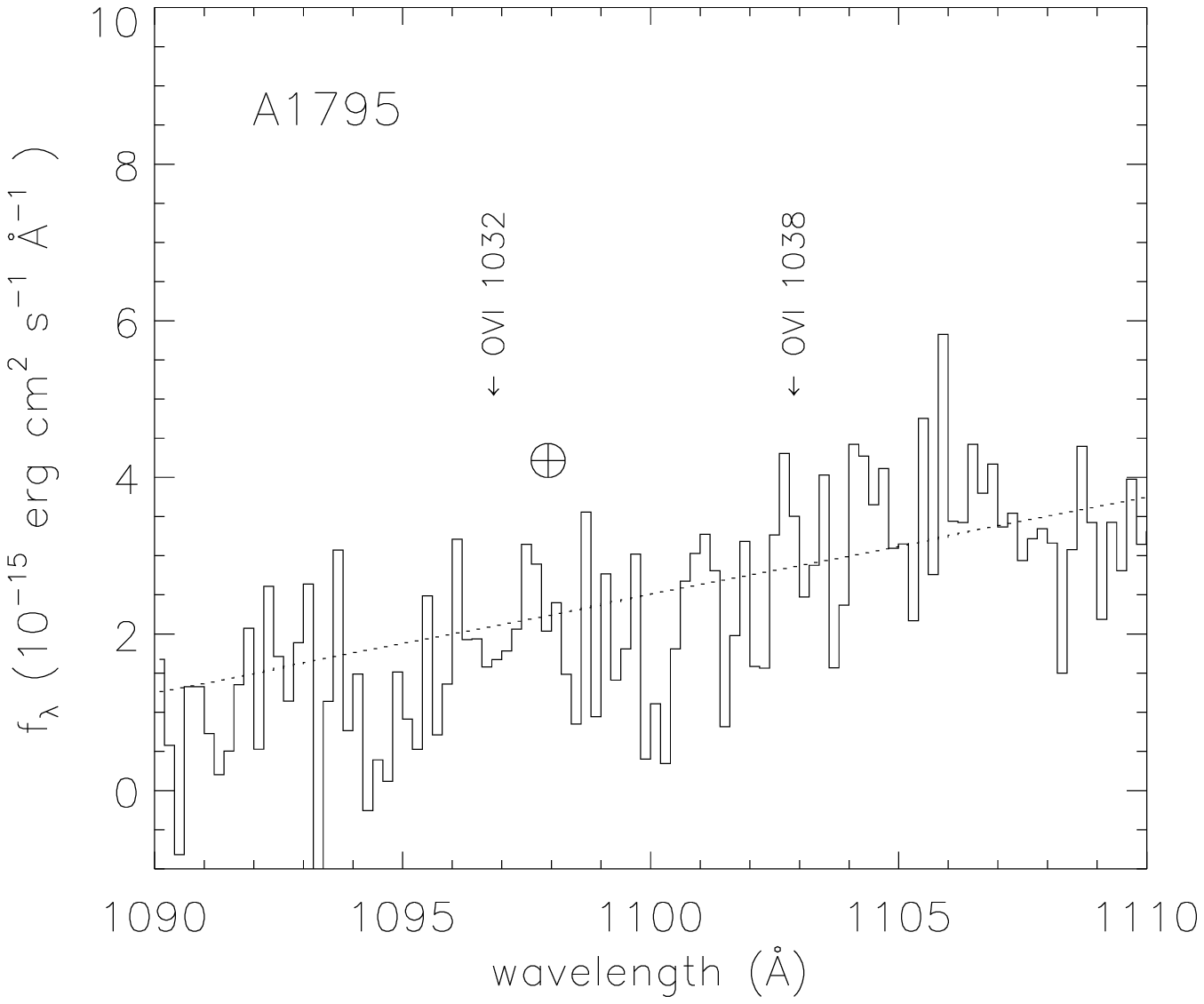}
\caption{A portion of the FUSE spectrum of A1795 taken during orbital
night (17 ksec) binned 
by 0.2\AA\/ ($\sim 60$ \kms\/). The spectral resolution for an 
extended source is $\sim 0.3$ \AA\/.
The expected positions of the redshifted \OVI\ \LL1032,1038 lines are
indicated.  A faint continuum from the cD galaxy is detected. 
Terrestrial airglow lines are indicated with an Earth ($\oplus$) symbol.
The dashed line in this figure, and subsequent figures, is the assumed
continuum.}
\end{figure*}

\newpage

\begin{figure*}
\plotone{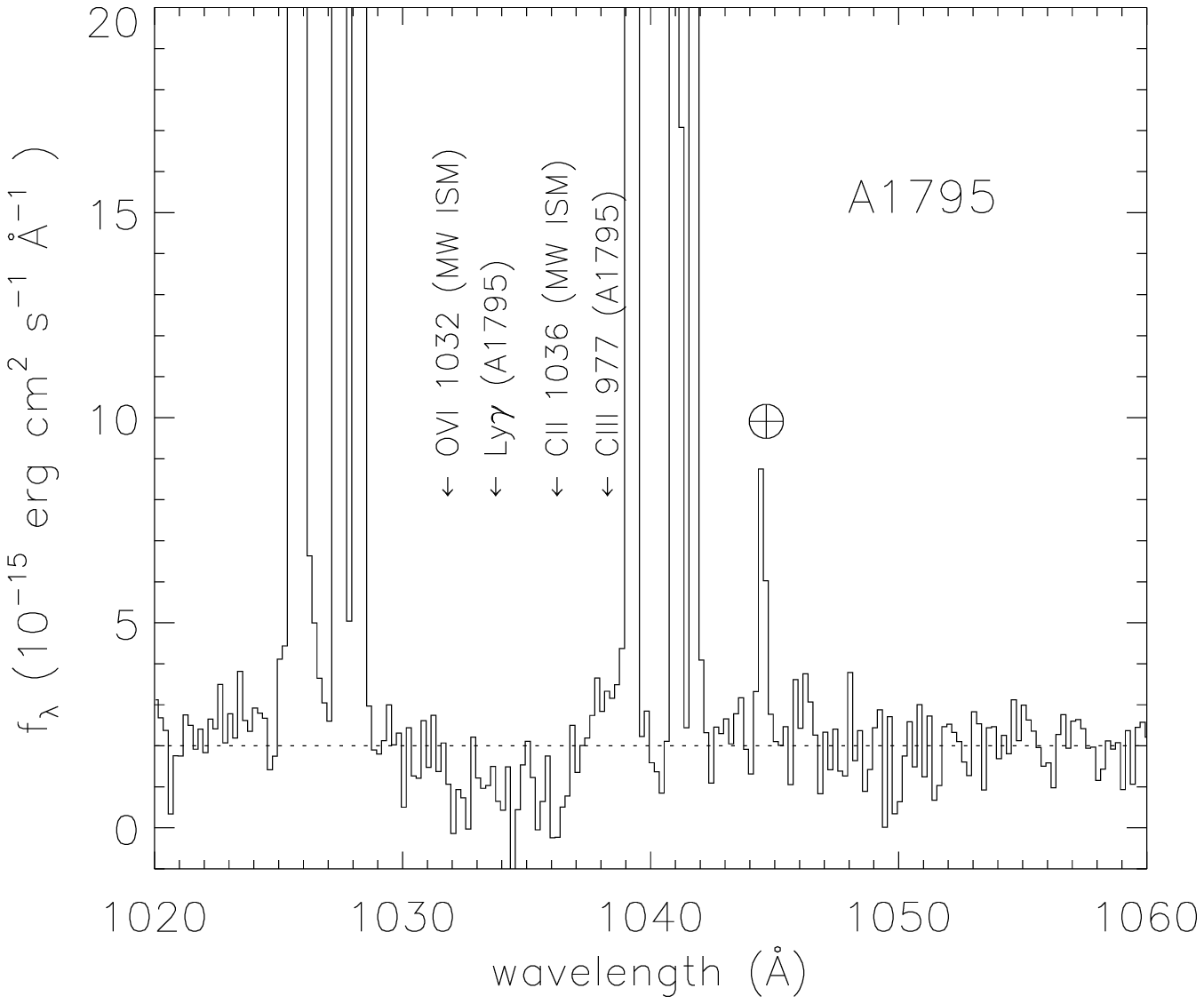}
\caption{A portion of the spectrum of A1795, roughly centered on redshifted
\CIII\ \lam977 (expected at 1038.67\AA\/). This spectrum was obtained
during orbital
day and night (37 ksec).  The data are binned
by 0.2\AA\/.  The strong emission line at 1025\AA\ is terrestrial Ly$\beta$,
and all the other strong emission lines at $\sim 1027$\AA\ and $1039-1041$\AA\
are airglow lines of \OI\/.  \CII\ \lam1036 and \OVI\ \lam1032 absorption
from the ISM in our Galaxy are indicated.  The expected position of 
redshifted \CIII\ \lam977 in A1795 is indicated.  }
\end{figure*}

\newpage 

\begin{figure*}
\plotone{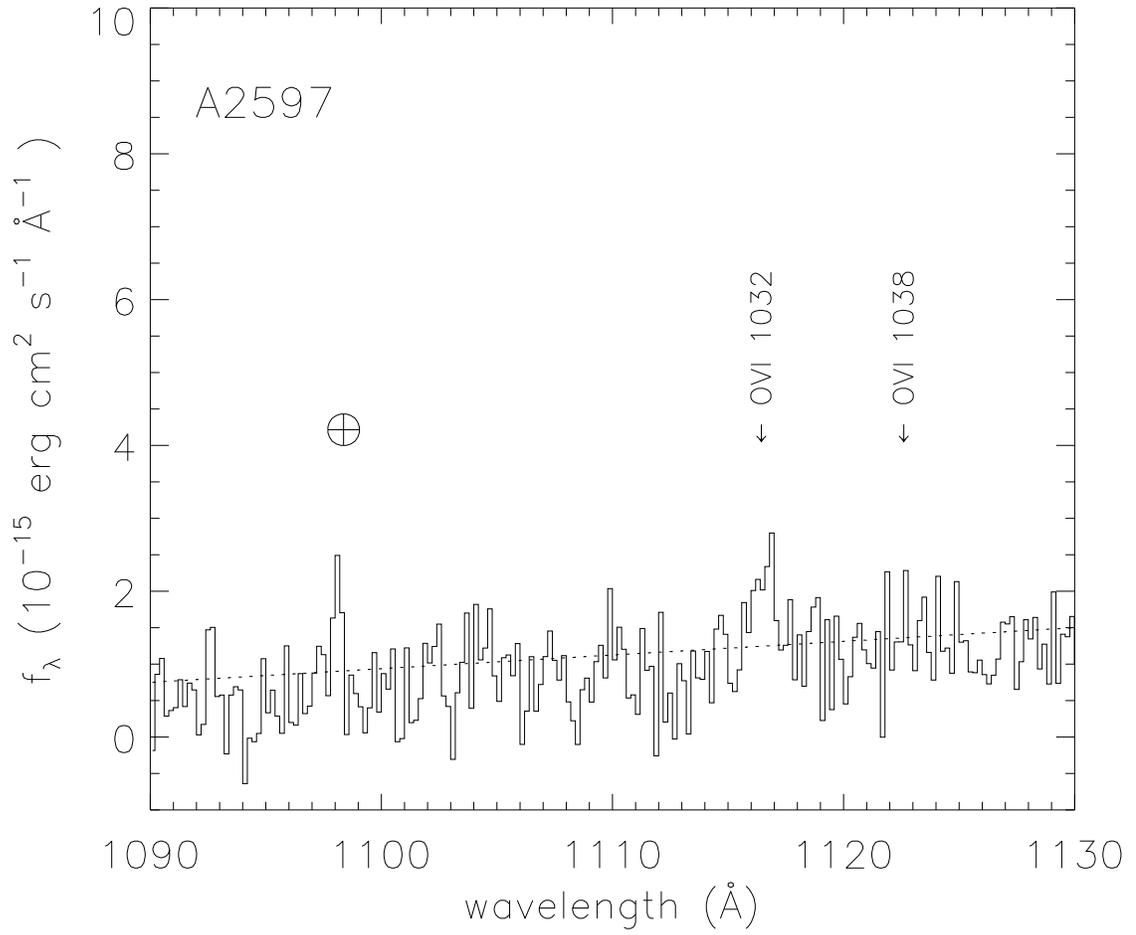}
\caption{A portion of the FUSE spectrum of A2597 binned
by 0.2\AA\/.
The expected positions of the redshifted \OVI\ \LL1032,1038 lines are
indicated.   }
\end{figure*}

\newpage

\begin{figure*}
\plotone{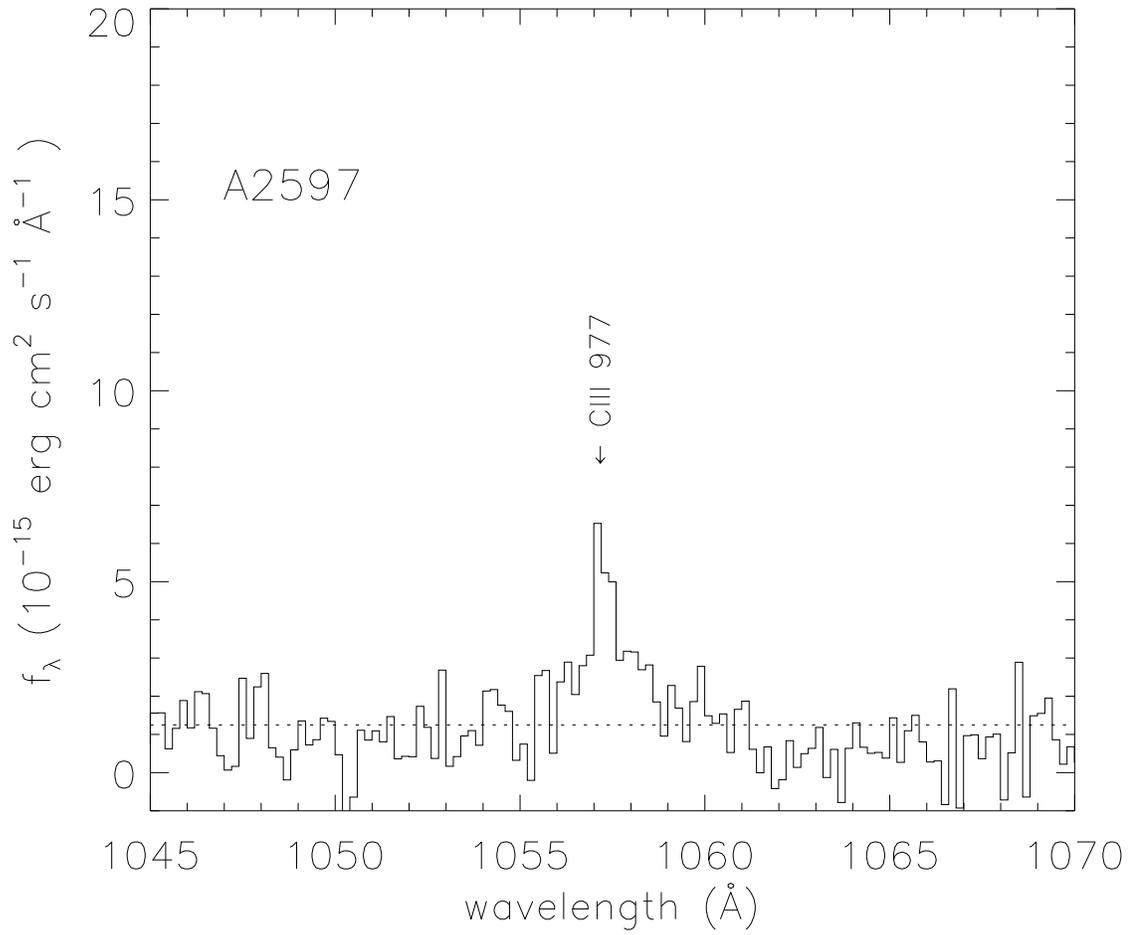}
\caption{A portion of the FUSE spectrum of A2597 binned
by 0.2\AA\/.  The expected position of the redshifted \CIII\ \lam977 line is
indicated.  }
\end{figure*}

\begin{figure*}
\plotone{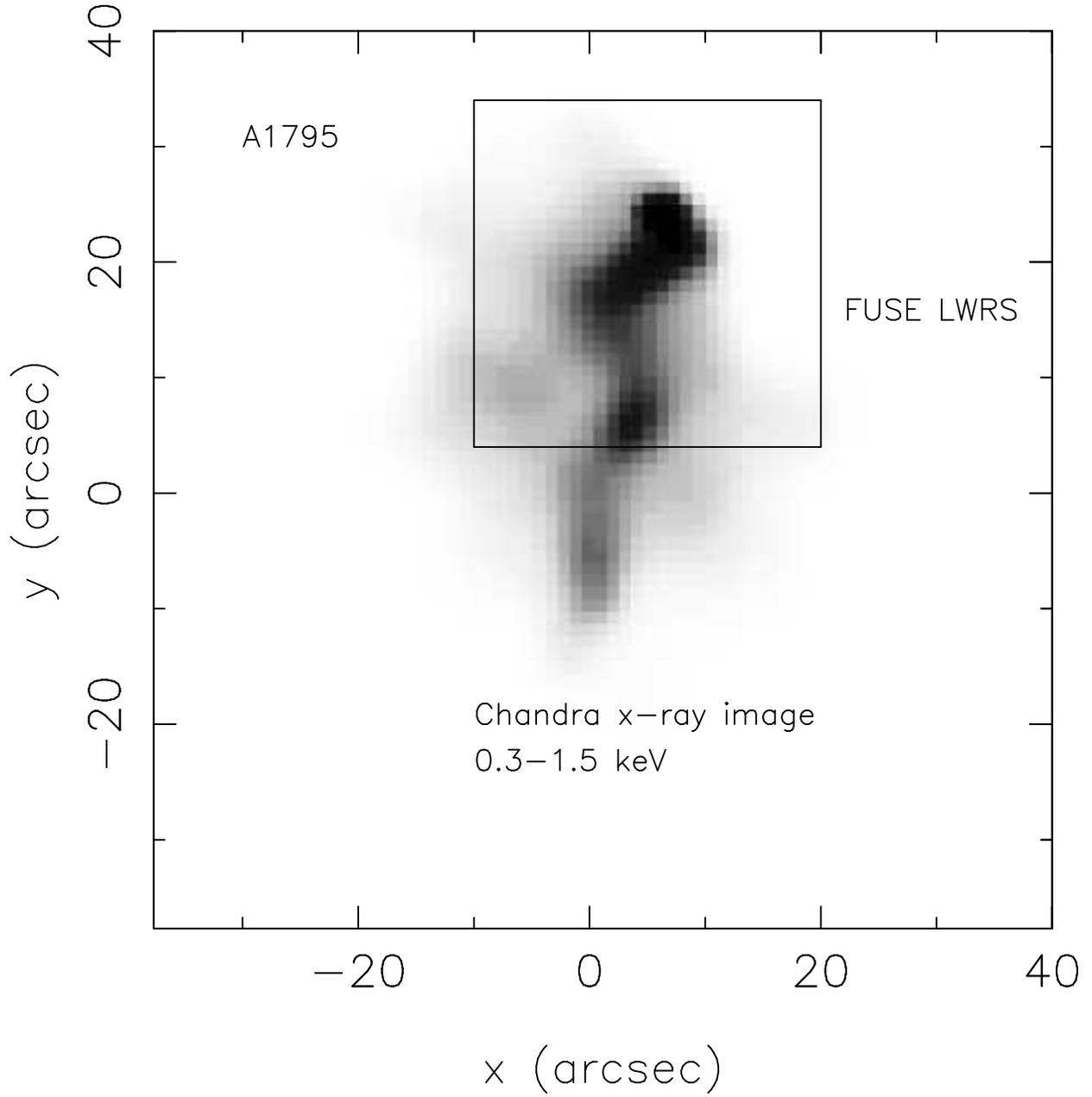}
\caption{A gray-scale image of the adaptively smoothed {\it Chandra} 
ACIS 0.3-1.5 keV
x-ray image from \citet{fabian01a} overlayed with the position of
the FUSE LWRS $30''\times30''$ spectrograph aperture.}
\end{figure*}

\end{document}